%% file: main.tex
\documentclass[nonacm, sigconf]{acmart}

\AtBeginDocument{%
  \providecommand\BibTeX{{%
    \normalfont B\kern-0.5em{\scshape i\kern-0.25em b}\kern-0.8em\TeX}}}

\copyrightyear{2026}
\acmYear{2026}
\setcopyright{acmcopyright}
\acmConference[CHI '26]{CHI Conference on Human Factors in Computing Systems}{April 13-17, 2026}{Barcelona, Spain}
\acmBooktitle{CHI Conference on Human Factors in Computing Systems (CHI '26), April 13-17, 2026, Barcelona, Spain}
\acmISBN{978-1-4503-XXXX-X/18/06}

\usepackage{lipsum}
\usepackage{subcaption}
\usepackage{macros}
\usepackage{float}
\usepackage{wrapfig}
\usepackage{cleveref}
\usepackage[dvipsnames]{xcolor}
\usepackage{mdframed}

\usepackage{csquotes}

\newmdenv[
  backgroundcolor=gray!15,
  leftmargin=4pt,
  rightmargin=4pt,
  skipabove=4pt,
  skipbelow=4pt,
  innerleftmargin=8pt,
  innerrightmargin=8pt,
  innertopmargin=6pt,
  innerbottommargin=6pt
]{takeaway}

\definecolor{RoyalBlue}{RGB}{26, 118, 198}
\definecolor{Lime}{RGB}{118, 185, 23}

\definecolor{hlblue}{HTML}{B5D7F4}
\definecolor{hlgreen}{HTML}{CFEBA5}
\definecolor{hlorange}{HTML}{F2BEB3}
\definecolor{hlfuchsia}{HTML}{CBC2FA}

\crefformat{section}{\S#2#1#3} 
\crefformat{subsection}{\S#2#1#3}
\crefformat{subsubsection}{\S#2#1#3}

\begin{document}

\title{Stepping off the Abstraction Ladder:\\A Survey and Conceptual Framework for\\Designing Abstractions in HCI}
\title{Beyond the Abstraction Ladder:\\A Survey and Conceptual Framework for\\Designing Abstractions in HCI}
\title{Beyond the Ladder:\\A Survey and Conceptual Framework for\\Designing Abstractions in Interactive Systems}
\title{How Do You Design Abstractions?:\\A Design Space and Conceptual Framework for\\Designing Abstractions in Interactive Systems}
\title{Abstractions In Design:\\A Design Space and Conceptual Framework for\\Designing Abstractions in Interactive Systems}
\title{Abstractions In Interfaces:\\A Design Space and Conceptual Framework for\\Designing Abstractions in Interactive Systems}
\title{Dimensions of Abstractions in Interactive Software Systems:\\A Design Space and Conceptual Framework}
\title{Making Abstraction Concrete:\\A Design Space and Conceptual Framework for\\Designing Abstractions in Interactive Systems}
\title{What is Abstraction?:\\A Design Space and Conceptual Framework for\\Designing Abstractions in Interactive Systems}
\title{Dimensions of Abstractions:\\A Design Space and Conceptual Framework for\\Designing Abstractions in Interactive Systems}
\title{What Do We Mean By Abstraction?:\\A Design Space and Conceptual Framework for\\Designing Abstractions in Interactive Systems}
\title{"What Do We Mean By Abstraction?":\\A Survey on How HCI Researchers Use Abstraction\\to Describe Designs of Interactive Systems}

\title{Dimensions of Interface Abstractions:\\A Design Space and Conceptual Framework for\\Designing Abstractions in Interactive Systems}
\title{"Does Abstraction Make Sense Now?":\\How HCI Researchers Use Abstraction to Describe the\\Designs of Interactive Systems}
\title{Making Sense of Abstraction:\\Terminology, Trends, and A Conceptual Framework for\\Designing Abstractions in Interactive Systems}
\title{"What Do We Mean By Abstraction?":\\A Design Space and Conceptual Framework for\\Designing Abstractions in Interactive Systems}
\title{Making Sense of Abstraction:\\Terminology, Design Space, and a Conceptual Framework for\\Designing Abstractions in Interactive Systems}
\title{Making Sense of Abstraction}
\title{What is Abstraction?}
\title{What Do We Mean by Abstraction?}
\title{Making Sense of Abstraction in Interactive Systems}
\title{Design of Abstractions in Interactive Systems}
\title{Abstraction Designs in Interactive Systems}
\title{How Are We Designing Abstractions in Interactive Systems?}
\title{Designing Abstractions in Interactive Systems}
\title{How Do We Design Abstractions?}
\title{Designing Abstractions in Interactive Systems:\\Terminology, Design Space, and Design Principles}
\title{How Do We Design Abstractions in Interactive Systems?:\\Terminology, Design Space, and Design Principles}
\title{How Do We Design Abstractions in Interactive Systems?}
\title[Making Sense of Abstraction]{Making Sense of Abstraction: How HCI Researchers Use Abstraction to Design Interactive Systems}
\title[Making Sense of Abstraction]{Making Sense of Abstraction: How the Term Abstraction Translates to Design Approaches of Interactive Systems}
\title[Making Sense of Abstraction]{Making Sense of Abstraction: How HCI Researchers Use and Interpret Abstraction to Design Interactive Systems}
\title[Making Sense of Abstraction]{Making Sense of Abstraction: How to Use and Interpret Abstraction to Design Interactive Systems in HCI}
\title[Making Sense of Abstraction]{Making Sense of Abstraction: How HCI Researchers\\Design Abstractions in Interactive Systems}
\title[Making Sense of Abstraction]{Making Sense of Abstraction: How HCI Researchers Design Abstractions in Interactive Systems}
\title[Making Sense of Abstraction]{Making Sense of Abstraction: A Survey on How HCI Researchers Design Abstractions in Interactive Systems}
\title{How Do We Talk About Abstraction?}
\title{How Do We Design Abstractions?}
\title{How We Design Abstractions}
\title{Designing Abstractions}
\title{What Do We Really Mean By Abstraction in Interactive Systems?}
\title{Designing Abstractions in Interactive Systems}
\title{What Do We Mean By Abstraction in Interactive Systems?}
\title{What Do We Really Mean By Abstraction?}
\title{Designing Abstractions in Interactive Systems Needs More Intentionality}
\title{Are We Deliberate Enough When Designing Abstractions in Interactive Systems?}
\title{Why Do We Need Abstractions in Interactive Systems?}
\title{What Do We Really Mean By Abstraction?}
\title{What is Abstraction?}
\title{Designing Abstractions}
\title{How We Design Abstractions}
\title{Interacting with Abstractions: An Interaction Model for Designing Abstractions in Interactive Systems}
\title{Interacting with Abstractions: An Interaction Model for Abstraction}
\title{Interacting with Abstractions}
\title{An Interaction Model for Abstraction}
\title[Gulfs and Abstraction]{Gulfs and Abstraction: An Interaction Model for Designing Abstractions in Interactive Systems}
\title[Abstraction Lens]{Abstraction Lens: An Interaction Model for Designing Abstractions in Interactive Systems}
\title[Abstraction Lenses]{Abstraction Lenses: An Interaction Model for Designing Abstractions in Interactive Systems}
\title[Abstraction as a Lens]{Abstraction as a Lens: An Interaction Model for Designing Abstractions in Interactive Systems}
\title{How Do We Design Abstractions?}
\title[Abstraction Spaces]{Abstraction Spaces: An Interaction Model for Designing Abstractions in Interactive Systems}
\title[Abstraction Arcs]{Abstraction Arcs: An Interaction Model for Designing Abstractions in Interactive Systems}
\title[Abstraction Arcs]{Abstraction Arcs: Reframing the Cognitive Process of Interaction Through the Lens of Abstraction}
\title[Abstraction Spaces]{Abstraction Spaces: Reframing the Cognitive Process of Interaction Through the Lens of Abstraction}
\title[Interaction Through the Lens of Abstraction]{Reframing the Cognitive Process of Interaction Through the Lens of Abstraction}
\title[Abstraction as a Lens]{Abstraction as a Lens to Reframe the Cognitive Process of Interaction}
\title[Abstraction Spaces]{Abstraction Spaces: Reframing the Cognitive Process of Interaction Through the Lens of Abstraction}
\title[Making Abstraction Concrete]{Making Abstraction Concrete: An Interaction Model of Abstraction in Interactive System Design}
\title[Making Abstraction Concrete]{Reframing the Cognitive Process of Interaction Through the Lens of Abstraction}
\title[Making Sense of Abstraction]{Making Sense of Abstraction: A Design Space and Interaction Model of Abstraction in Interactive Systems}
\title[Making Abstraction Concrete]{Making Abstraction Concrete: A Design Space and Interaction Model of Abstraction in Interactive Systems}
\title[Making Abstraction Concrete]{Making Abstraction Concrete: How Interactive Systems Encode The Process of Abstraction}
\title[Making Abstraction Concrete]{Making Abstraction Concrete: How Interactive Systems Perform the Work of Abstraction}
\title[Making Us Work in Abstraction]{Making Us Work in Abstraction: How Interactive Systems Implement the Process of Abstraction to Control and Empower Users}
\title[Making Abstraction Work]{Making Abstraction Work: Characterizing the Interaction Dynamic Between Users and Interactive Systems as Abstraction Work}
\title[Abstraction as Work]{Abstraction as Work: Characterizing the Interaction Dynamic Between Users and Interactive Systems Through the Lens of Abstraction}
\title[Making Abstraction Concrete]{Making Abstraction Concrete: }
\title[Beyond the Ladder]{Beyond the Ladder: When Developing Abstractions in Interactive Systems}
\title[Abstraction Goes Beyond the Ladder]{Abstraction Goes Beyond the Ladder When Developing Interactive Systems}
\title[Beyond the Ladder]{Beyond the Ladder: How HCI Uses the Principle of Abstraction to Design Dynamic Interactive Systems}
\title[Beyond the Ladder]{Beyond the Ladder: How Do We Use the Principle of Abstraction to Build Dynamic Interactive Systems?}
\title[Beyond the Ladder]{Beyond the Ladder: Using the Principle of Abstraction to Build Dynamic Interactive Systems}
\title[Making Abstraction Concrete]{Making Abstraction Concrete: A Design Space and Interaction Model of Abstraction in Interactive Systems}


\author{Bryan Min}
\email{bdmin@ucsd.edu}
\affiliation{%
  \institution{University of California San Diego}
  \streetaddress{9500 Gilman Dr}
  \city{La Jolla}
  \state{California}
  \country{USA}
  \postcode{92093}
}

\author{Sangho Suh}
\email{sanghos@allenai.org}
\affiliation{%
  \institution{Allen Institute for AI}
  \streetaddress{3800 Latona Ave NE, Suite 300}
  \city{Seattle}
  \state{Washington}
  \country{USA}
  \postcode{98105}
}

\author{Jim Hollan}
\email{hollan@ucsd.edu}
\affiliation{%
  \institution{University of California San Diego}
  \streetaddress{9500 Gilman Dr}
  \city{La Jolla}
  \state{California}
  \country{USA}
  \postcode{92093}
}

\author{Haijun Xia}
\email{haijunxia@ucsd.edu}
\affiliation{%
  \institution{University of California San Diego}
  \streetaddress{9500 Gilman Dr}
  \city{La Jolla}
  \state{California}
  \country{USA}
  \postcode{92093}
}

\renewcommand{\shortauthors}{Min et al.}

\begin{abstract}
The principle of abstraction guides the design of interactive systems, yet we lack a conceptual framework to understand how it shapes interaction design. Existing models, such as the gulfs of execution and evaluation, do not explicitly model abstractions in the system or in users’ mental models, and therefore lack actionable guidance for designing abstractions. To investigate how abstractions are employed in interactive systems, we surveyed 457 papers and synthesized a design space of abstraction techniques along six dimensions. We use this design space to reframe the gulfs through a lens of abstraction, explicitly articulate the cognitive and design processes by which users and systems bridge and navigate the abstraction gap, and demonstrate how this model integrates existing perspectives and surfaces new opportunities for future systems.
\end{abstract}



\begin{teaserfigure}
    \includegraphics[trim=0cm 0cm 0cm 0cm, clip=true, width=\textwidth]{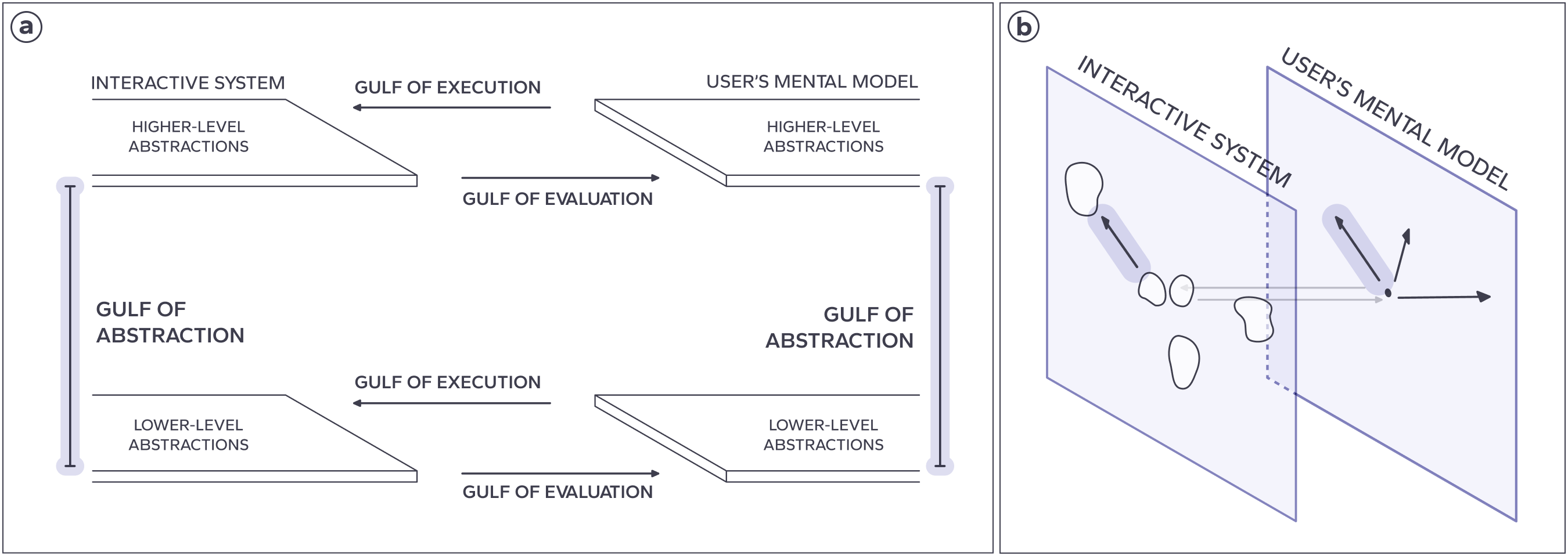}
    \caption{We propose Abstraction Spaces, an interaction model of abstraction. Abstraction Spaces describes how abstractions formed in both the user’s mental model and the interactive system influence the interaction dynamic between them. (a) We introduce the Gulf of Abstraction, which denotes the distance between the abstractions in the user’s mental model and those in the system, making misalignments explicit. (b) Zooming out, our model highlights that abstraction in users and systems is not strictly linear but instead spans a broader space that both can navigate.}
    \label{fig:teaser}
    \Description{Conceptual model of the Gulf of Abstraction. (a) Extension of Norman’s gulfs of execution and evaluation with a vertical dimension for higher- and lower-level abstractions. (b) 3D schematic showing the interactive system and user’s mental model as parallel planes, with arrows illustrating how actions span and align across abstraction levels.}
\end{teaserfigure}

    \maketitle{}

\input{sections/1_introduction}
\input{sections/2_background}
\input{sections/3_definition}


\input{sections/5_design_space}

\input{sections/6_abstraction_model}
\input{sections/9_discussion}

\input{sections/99_conclusion}


\bibliographystyle{ACM-Reference-Format}
\bibliography{main, survey}

\appendix
\input{sections/999_appendix}

\end{document}

%% file: sections/1_introduction.tex
\section{Introduction}
\label{section:introduction}

Abstraction, understood as both a process of focusing on desired details of information and an outcome of modeling information, shapes every facet of interaction.
Interactive systems employ abstractions in both the interface content and operations to facilitate how users understand and interact with these systems.
It is striking, however, that HCI lacks an interaction model that explicitly accounts for how abstraction affects the cognitive process of interaction, given its foundational role.
The canonical model of the Gulfs of Execution and Evaluation specifies the key stages of action bridging the gulfs between the users and systems, but only addresses abstraction indirectly by acknowledging that the abstractions employed in system can result in rough changes in distances of the gulfs \cite{hutchins1986directmanipulation}. The absence of the deliberate treatment of abstraction leaves little actionable guidance for designing effective abstractions or for holistically understanding the cognitive processes revolving around abstractions.

This is understandable, as the Gulfs model was developed during the paradigm shift toward direct manipulation interfaces four decades ago, and therefore, emphasized low-level manipulative aspects of interfaces. In recent years, the research community has observed a shift in abstraction where an increasing number of works are exploring the design of dynamic and high-level abstractions for interaction. Additionally, we are at a new paradigm shift of how computational systems model the world with the recent AI breakthroughs of Large Language Models which enable new forms of interaction between human and AI-powered systems. We believe it is time to revisit the classic model of the cognitive process of interaction through the lens of abstraction. Therefore, our research seeks to answer the following questions.  First, what abstraction techniques and perspectives has the HCI community recently developed concerning interactive systems? Second, to what extent can these perspectives and techniques be captured within the Gulfs model, and where do they impose pressures on it? Third, if reform of the existing model is warranted, what changes are required? 

We surveyed 457 papers across major HCI venues to understand the landscape of abstraction techniques in interactive systems, and synthesized a design space spanning six dimensions. These dimensions capture the contexts in which abstractions are constructed, the types of transformation techniques employed, and the ways abstractions are manifested in interfaces. Examining these techniques against the traditional Gulfs model reveals its three key limitations.

First, although the Gulfs model describes the stages pertinent to abstraction within users' mental models, such as intention formation, action specification, and interpretation, it omits abstraction transformations that occur within \textit{systems} and how these various transformations influence the user's mental model. Second, our survey analysis reveals that abstraction influences all stages of the gulfs. Therefore, to better understand the effects of abstraction on the entire cognitive and computational process of interaction, it may be more productive to situate the gulfs themselves within the broader context of abstraction, rather than describing how abstraction plays a role in each stage. Third, our survey surfaces various ways designers have been incorporating diverse abstractions to support users in their tasks by considering multiple levels of abstraction, yet research has been showing even multiple levels are not enough to support their diverse information needs \cite{min2025malleable, williams2023data}. Instead, it may be more fruitful to conceive of the cognitive process as operating within a \textit{space} of abstractions.


With these findings, we propose a reframing of the traditional Gulfs model through \textit{Abstraction Spaces}, an interaction model of abstraction. Abstraction Spaces adds three key components. First, we introduce the Gulf of Abstraction, which describes the distance between the abstractions employed in the user’s mental model and those provided by the system. This gulf is represented as a vertical dimension of misalignment. Second, we distinguish between two types of abstraction misalignments: undershoots and overshoots. Undershoots occur when the system provides a level of abstraction that is too low for the user’s needs, while overshoots occur when the system provides a level that is too high. Finally, we illustrate how designing with abstraction involves a \textit{space} of abstractions within the interaction model, encouraging designers to view abstraction not as a single layer or ``ladder'' \cite{victor2011ladder}, but as a broad design space.


We examine how Abstraction Spaces can be used to model existing interaction workflows and perspectives, and to surface design opportunities within them. Through this analysis, we show how our interaction model highlights the reciprocal influence between system and user abstractions, identifies the boundaries of what can be modeled, and reveals opportunities for designing systems that not only align with the user’s mental model but also deliberately misalign---encouraging users to realign their understanding toward the system's abstractions.

%% file: sections/2_background.tex
\section{Related Work}
\label{section:background}
We first ground our work within the broad scope of research on abstraction, and then review a small set interactive systems that employed multiple abstractions to paint a rough picture of the design of abstractions in HCI. A more comprehensive account of various abstraction design techniques can be found in section 4; we then discuss the connections between notable interaction models and frameworks with abstractions.

\subsection{The Power of Abstraction}
The concept of abstraction refers to both the process of forming a new scheme of information, and the scheme itself. The purpose of abstraction, as pointed out by Dijkstra, is ``not to be vague, but to create a new semantic level in which one can be absolutely precise'' in depicting a problem or phenomenon \cite{dijkstra1972humble}. As a powerful human cognitive ability that serves as the foundation of intelligence, abstraction has been widely explored in different research fields \cite{newell1994unified, minsky1986society, engelbart2023augmenting, bruner1966concretenessfading, hayakawa1967language}. For example, abstraction has been regarded as a core skill for programming and computational thinking \cite{wing2006computational, dijkstra1972humble}.
Research in linguistics, visual and spatial cognition has found that people fluidly shift the levels of abstraction to communicate a concept via natural language expressions \cite {hayakawa1967language} as well as visual or gestural depictions \cite{tversky2019mind}. In the field of semiotics and visualization, encoding information and data using visual symbols has been seen as a process of forming a visual abstraction of information \cite{viola2017pondering}. Utilizing the gradual transition of a concrete depiction of a problem to a more abstract one to facilitate knowledge understanding, also known as concreteness fading, has been established as an effective pedagogical methodology \cite{bruner1974toward, fyfe2019making}. Research in artificial intelligence explores invoking the abstraction process in LLMs to increase their performance on various reasoning tasks \cite{zheng2023take}. In diverse engineering domains, the deliberate use of abstraction to define higher-level, reusable modules constitutes a foundational strategy for managing complexity in system design—spanning, but not limited to, programming languages \cite{cardelli1985understanding}, software and hardware systems \cite{gamma1995elements}, architectures \cite{alexander1977pattern}.

Given the profound influence of abstraction on both human cognitive processes and the system computational processes, our research seeks to enable HCI scholars to more explicitly and systematically examine the role of abstraction across the holistic process of human-computer interaction.

\subsection{Abstraction in Interactive Systems}
The concept of abstraction has been deeply employed in interactive systems across various interaction paradigms. Early terminal-based interactive systems incorporate the concept of abstraction by representing low-level system calls into compact commands. Sketchpad, the first graphical user interface, represents abstractions, such as constraints of geometric shapes, as graphical manipulable elements to enable direct manipulation of mathematical relationships \cite{sutherland1963sketchpad}. Natural language interfaces enable users to express their intents to the system without resorting to low-level programs \cite{winograd1972understanding} and manual controls \cite{bolt1980put}. Recent advances in AI models drastically expand the types of content that can be generated with intents described at a higher-level of abstraction (e.g., merely specifying a goal) \cite{bubeck2023sparks}.

One notable design pattern emerged from the past decades is the employment of multiple abstractions in interactive systems. A canonical example is semantic zooming, which transforms the content to different levels of abstraction according to the zooming level \cite{bederson1994pad}, which is widely used in later systems \cite{suh2023sensecape, suh2024luminate, yoon2015semantic}.
Additionally, Victor popularized the use of the ``ladder of abstraction'' as a metaphor and design approach to design multiple levels of abstraction in interactive systems, encouraging designers to enable users to move between levels of abstraction as a powerful way to gain insights into a system \cite{victor2011ladder}.
Xia et al.'s WritLarge systems explored representing hand-drawn graphics at different levels of abstraction, including, for example, formal digitization of pen strokes enable users to interact with the desired level of abstraction, from pixels to table components, accessed by with system provided representational axes \cite{xia2017writlarge}. Min et al. recently explored supporting malleable abstractions for overview-detail interfaces, enabling end-users to flexibly form their desired level of abstraction \cite{min2025malleable}.
This pattern of multiple abstraction has also been broadly applied to other domains such as data visualization \cite{halladjian2020scale, halladjian2022multiscale}, programming visualization \cite{hayatpur2023crosscode, hayatpur2025shapes}, as well as LLM-driven interfaces \cite{liu2023what}. 


One of our aims is to uncover the abstraction design techniques that HCI and adjacent communities have developed, building a shared collection of methods that can be reused and extended. More importantly, we articulate a design space of abstraction defined by key dimensions that can guide designers in systematically creating, evaluating, and extending new abstraction strategies.

\subsection{Abstraction In Interaction Models}
Despite the proliferation of systems employing a diverse array of techniques for designing abstractions, the HCI field lacks an interaction model that explicitly theorizes and formalizes abstraction. Nonetheless, traces of such a model can be discerned within several existing frameworks. 

The Instrumental Interaction model proposes three design principles for design interactive systems and functionality: reification, polymorphism, reuse \cite{mbl2000instrumentalinteraction, instrumentalavi}. While these principles directly build atop the concept of abstraction, they do not model the abstractions employed in interactive systems. The sensemaking model incorporate abstraction by depicting a learning loop where users iteratively form new abstractions of information \cite{russellsensemaking}. This model is similar to the Gulfs in that they both focus on describing the process of interaction, but they do not model the relationship between abstractions formed in users' mental models and abstractions designed in systems. The Gulfs, developed initially and partially, as a response to Shneiderman's high-level characterization of direct manipulation interfaces \cite{shneiderman1983direct}, focused on the nuanced dissection of factors that could increase or decrease the feeling of directness. While it acknowledges that high-level abstraction employed in the interfaces could shorten both the gulfs, the framework does not explicitly model abstractions in interactive systems \cite{hutchins1986directmanipulation}.

We are motivated by the lack an interaction model that directly accounts for abstraction. In this work, we adapt the framework of the Gulfs, due to its ability to generally characterize the cognitive process of interaction with diverse systems, and augment it with a new Gulf of Abstraction.
By examining a diverse set of abstraction design techniques and comparing with existing models, we aim to examine how our new model demonstrates analytical, critical, and constructive power \cite{beaudouinlafon2021generative}.

%% file: sections/3_definition.tex
\section{What is Abstraction?}
Abstraction is a challenging concept to define, as any definition that proves effective in one context may be deemed inadequate in another. Nevertheless, we offer our working definition to help readers understand the present work and to provide a shared vocabulary that the community can collectively evaluate and refine by using it to describe the cognitive and computational processes of human–computer interaction.


\textbf{Abstraction} denotes a specific schema of information, depicting the structural configuration of entities and their relations. For example, a digital photo album typically embodies two system abstractions: one schema that organizes all the photos in the album (e.g., by date, tags, or albums), and another schema that specifies the pixel-based structure of each individual image. Similarly, a user interacting with this system may form two corresponding mental abstractions: one that conceptualizes how the collection of photos is organized as a whole, and another that focuses on how the images can be manipulated at pixel level such as adding filters or removing regions of pixels. Together, these system and mental abstractions, shape how information is stored, retrieved, and experienced.

Our definition of abstraction departs from prevailing accounts in two important respects. First, we deliberately deemphasize the common association of abstraction with “higher-levelness,” simplicity, or opposition to concreteness. We argue that whether an abstraction appears “high-level” or “simple” is a relative notion, heavily shaped by users’ conceptual interests and their familiarity with the domain. For instance, while program code may be considered an abstract syntactic notation relative to run-time values—or appear abstract to novice programmers—it nonetheless specifies computational procedures concretely and unambiguously and can be intuitively manipulated by proficient programmers. Second, our definition concerns the schema itself rather than the process that generates it. This distinction is important because it is notoriously difficult to disentangle the process of abstraction from notions of simplification or generalization. For instance, it is semantically awkward to describe “abstracting” a piece of content into a more concrete form. By treating abstraction as the schema rather than the act, we can coherently refer to both abstract and concrete instances as specific abstractions.

Abstraction and representation are two concepts that are often conflated in literature, as they are semantically and functionally intertwined. Here we provide a working definition of representation.
 
\textbf{Representation} denotes the concrete instantiation of such an abstraction in a specific medium, notation, or format that renders the information perceptible and operable. With the example of photo album, the abstractions are manifested in the form of, for example, a grid based visual representation of all the images, and a single view representation of an individual image with corresponding pixel-based editing functionality.

With our definitions, every representation presupposes an underlying abstraction, yet multiple representations may embody the same abstraction across different modalities (for example, the same message expressed textually or auditorily). While our focus is on abstraction, as shown in the design space later, many abstraction design techniques we identified are centered on the representations of the abstractions to enable users to interact with the abstractions, such as the amount of detail to present given a specific abstraction and representation, how different representations of the same or different abstractions should be presented to the users.

%% file: sections/5_design_space.tex
\section{Design Space of Abstractions in Interactive Systems}
\label{section:design_space}

We grounded the development of our interaction model through a survey of interactive systems that design abstractions in HCI.

\subsection{Methodology}

\paragraph{\textbf{Gathering Relevant Publications}}

We conducted a systematic literature search of three research databases, ACM Digital Library, IEEE Xplore, and Google Scholar using the terms ``abstraction,'' ``abstracting,'' ``abstracted,'' and ``abstractive'' across seven HCI venues (CHI, UIST, DIS, IUI, TOCHI, PACMHCI, TIIS) and four HCI-related venues, accessibility (ASSETS), programming interfaces (VL/HCC) and visualization (TVCG, EuroVIS).
We excluded less rigorously reviewed publications such as extended abstracts, demos, posters, and workshop papers.
After merging duplicates, we gathered a total of 3128 papers.

\paragraph{\textbf{Filtering}}

Our primary inclusion criterion is \textit{whether the paper contributes an interactive system and uses the principle of abstraction as a design approach}. 
To build the collection of papers that match this criterion, we excluded empirical studies, surveys, and technical/performance contributions. We also excluded framework papers that did not directly contribute design approaches for interactive systems, such as frameworks for meta-research or technical frameworks for software engineering.
We also excluded papers that used the term abstraction in an irrelevant context. This included cases such as describing the phenomena of real-life activity unrelated to human-computer interaction, describing the paper's analysis or literature review task, or describing a feature of a different paper's system. We further detail our exclusion criteria in Appendix \ref{appendix:criteria}.
After filtering, we finalized a collection of 457 papers.

\paragraph{\textbf{Tagging}}

We iteratively developed a codebook from our collection of papers by conducting thematic analysis \cite{braud2006thematicanalysis}. 
Our codebook consisted of two classes of codes: (1) how paper authors used the term abstraction to describe their design approaches and (2) what were the resulting interface design elements that utilize abstraction.
We began with an initial set of tags based on how ``abstraction'' was used in the paper, the purpose of the research system, and the design components of the interactive system.
Throughout our tagging activities, we frequently discussed emerging patterns and additional dimensions to surface.

\paragraph{\textbf{Limitations}}

Since there is at least some aspect of abstraction designed into interactive systems, it is almost impossible to comprehensively survey the entire design space.
Our goal is not to provide an exhaustive survey, but to sample a broad design space of abstraction design techniques.

\subsection{Design Goals of Abstraction}
\label{section:design_goals}

\begin{figure}
    \centering
    \includegraphics[width=1\linewidth]{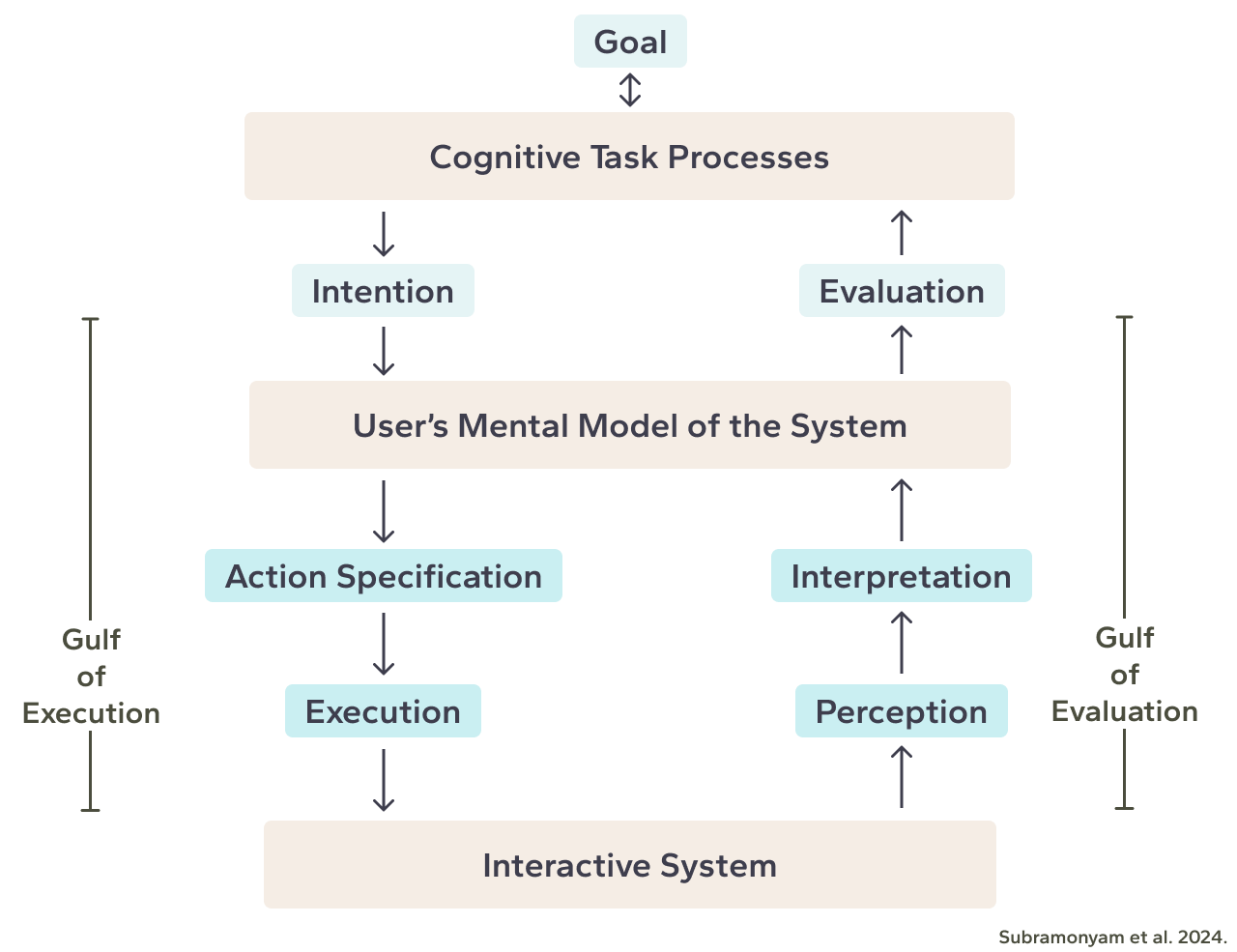}
    \caption{Subramonyam et al.'s Expanded View of Norman's seven-stage model of interaction \cite{hari2024gulfenvisioning} adds cognitive task processes and system mental model blocks to the traditional seven-stage model. This model highlights four stages between the system mental model and the interactive system.}
    \label{fig:design-purpose-stages}
    \Description{Extended model of Norman’s seven stages of action, showing the user’s goals, cognitive task processes, mental model, and interaction with the system across the Gulfs of Execution and Evaluation. }
\end{figure}

Our first set of codes examined how authors described their system's purpose for abstraction. In summary, these purposes align with four of Norman’s seven stages of action \cite{norman1988psychology}: perception, interpretation, action specification, and execution. We draw this boundary from Subramonyam et al.'s expanded view of Norman's framework \cite{hari2024gulfenvisioning}, which distinguishes between stages that involve the user’s interaction with the system and those that involve the user’s planning and evaluation. Specifically, the four stages above correspond to spanning the gulfs of execution and evaluation, while the remaining three stages---forming a goal, forming an intention, and evaluating outputs---capture the user’s cognitive process of comparing the system's output with their goals and determining their plans for future interaction (Fig. \ref{fig:design-purpose-stages}).

\paragraph{\textbf{Perception}}


Designing abstractions in the interface to influence how users perceive information can stem from two primary purposes. First, abstractions can reduce clutter in the interface, making information visually cleaner. This is often done to streamline workflows in complex systems such as information visualization systems \cite{mayorga2013splatterplots} and sensemaking tools \cite{suh2023sensecape}. By reducing clutter, interfaces also help users understand and manage information more effectively. 
Second, abstractions can help orient users as they navigate the interface. For example, the system may provide abstractions of a large corpus of text documents \cite{kang2023synergi} or of a spatial canvas of media \cite{koch2020semanticcollage}. These abstractions improve the user’s awareness of their position within an interaction task and improve navigation.



\paragraph{\textbf{Interpretation}}

Abstraction designs also aim to influence how users interpret the information displayed in the interface. This can include supporting users in forming generalizations of data \cite{weinman2021improving, kang2025biospark}, recognizing patterns \cite{wallner2019aggregated}, or creating classifications \cite{zimmerman2009designing}. Abstractions can also encourage users to adopt new perspectives on the information.
For example, a system may present a color space of a photo, enabling users to evaluate the aesthetics of images through the surfaced color properties \cite{beaudouinlafon2023color}.


\paragraph{\textbf{Action Specification}}


Designing abstractions also aims to support how users translate their intents into concrete actions within the interactive system. This may involve helping users express their intent more naturally through intuitive gestures \cite{schofield2020magical} or recognizing abstract intents spoken by the user and translating them into concrete operations for the system \cite{tilekbay2024expressedit}.


\paragraph{\textbf{Execution}}


Lastly, designing abstractions aims to help users execute their specified actions. This can include tools for programming-by-demonstration, where users automate repetitive tasks \cite{li2017sugilite, pu2023dilogics}, or techniques that encapsulate multiple actions into common operations, such as automatically zooming out as a user scrolls more quickly through a document \cite{igarashi2000speeddependent}.




\subsection{Design Dimensions of Abstraction}



Researchers design abstractions to achieve their design goals through various design approaches, and these approaches take shape with diverse design elements in the interface. For instance, if a design goal is to make a text document ``easier to understand,'' a researcher might (1) reduce its complexity by summarizing key points, (2) change its representation by visualizing sentences as interactive figures, or (3) extract commonalities across related documents the user is already familiar with.

In the literature, all of these design approaches are often described as ``forming an abstraction'' of the document, yet they are fundamentally different: one simplifies, another provides more concrete visuals, and another broadens the user's view. In this section, we aim to unpack these differences by identifying the different dimensions of design elements researchers implement when designing abstractions.
We identify six such dimensions:


\begin{enumerate}
    \item \textbf{Unit of Interest}: \textit{What does the system form abstractions of?}
    \item \textbf{Granularity}: \textit{What types and amount of detail does the transformation provide?}
    \item \textbf{Representation}: \textit{How does the abstraction take form in the software interface?}
    \item \textbf{Transformability}: \textit{How much can users modify the abstractions?}
    \item \textbf{Presentation}: \textit{How does the software interface situate and display abstractions?} 
    \item \textbf{Guidance}: \textit{How does the interface influence and scaffold the user’s workflow with the provided abstractions?}
\end{enumerate}



\subsubsection{\textbf{Dimension 1: Unit of Interest}} \textit{What element does the system form abstractions of?}
\label{section:unit_of_interest}



Abstraction involves transforming one element of information into another. This dimension describes what the unit of this element is. We categorize two types of elements that is determined as a particular unit: content and operations.



\paragraph{\textbf{Content}}


First, the unit of abstraction can be the content in the interface.
The types of content can be individual data points in a plot \cite{cuenca2018multistream, cakmak2020motionglyphs}, nodes in a diagram \cite{calo2025deepflow, sarkar1993stretching}, or functions or lines in code \cite{louie2022affinder}.
The unit of abstraction is the content that is of interest and can range in scope. For example, the unit in a text document can be considered as the individual text characters \cite{jiang20231dtouch}, paper sections \cite{palani2023relatedly}, or even the entire document itself \cite{chen2023synergy}.
Some systems use the document as a view to contain and represent abstractions of its text context \cite{masson2023charagraph, gobert2022ilatex}, while others use the document as the object to present abstractions over many different documents to aid understanding and synthesis \cite{koch2014varifocalreader, palani2023relatedly}. Similarly, the unit of interest in a sketching canvas can be considered the individual strokes \cite{wallner2019aggregated} or the shapes formed by the sketches \cite{saquib2021constructing}, while the unit of interest in a video can be considered the video as the media object itself \cite{mrth2023scrollyvis, lu2024knotation} or the individual frames making up the video \cite{matejka2013swifter, cheng2009smartplayer}.
The unit of interest can also expand to the scope of the space or view of information itself. For example, users can stretch \cite{sarkar1993stretching}, fold \cite{elmqvist2010mlange}, or distort \cite{pindat2012jellylens} the space of a map to better focus on key areas.

\paragraph{\textbf{Operations}}


Another set of elements that can be transformed through abstraction is operations.
Identifying the unit of abstraction for operations may involve the unit of content to be selected—for example, from concrete times (e.g., 7 am) to higher-level times (e.g., morning) \cite{heer2008generalized}, the unit of program functions used to author different forms of media \cite{hempel2016semiautomated}, the unit of mouse scrolls for navigating an interface \cite{igarashi2000speeddependent}, or the unit of text selection, ranging from individual characters to phrases \cite{jiang20231dtouch}.
Operations can also be abstracted at the level of user expressions. These expressions take the form of a range of modalities, such as direct manipulation, sketches, gestures, typing, or speech \cite{ma2022layered, peng2024designprompt, saquib2019interactive, scaffidi2009intelligently, kim2022stylette}. Through these modalities, users instruct the system in different ways. For instance, some speech-to-text systems enforce concrete units of spoken expression, such as speaking directly in Java code \cite{begelspoken}, while others support ambiguity by relying on large language models \cite{tilekbay2024expressedit}.

\subsubsection{\textbf{Dimension 2: Granularity}}\textit{What types and amount of detail does the transformation provide?}
\label{section:granularity}

This dimension concerns how much detail is involved when presenting content in the interface. Researchers has explored both adding and reducing complexity, as well as managing specificity of elements. What constitutes as details are determined by the employed abstraction. For text, images, and videos, the common abstractions model the details as words, pixels, and frames.
Much of other types of digital content can be modeled with a wide range of data formats that emphasize different aspects of the information. For example, a common abstraction of digital content is to model it by attributes, which makes the amount of detail of content more quantifiable and manipulable,

\paragraph{\textbf{Removing Complexity}}

Significant amount of research explored reducing the amount of information users have to process and help them understand its key ideas more efficiently. For example, interfaces can provide summarizations and descriptions of text \cite{suh2023sensecape, jiang2023graphologue, yuan2023critrainer, boydell2007social}, images \cite{hammond2010creating, l2021dynamic, shi2023destijl, peng2024designprompt, koch2020semanticcollage}, videos \cite{pongnumkul2010contentaware, ma2022layered, mrth2023scrollyvis, matejka2013swifter, lu2024knotation}, user activities \cite{zhang2025summact}, and code \cite{liu2023what, yan2024ivie}. 
When the content employs an attribute-based abstraction, details can be removed by removing the attributes completely \cite{min2025malleable} or simplifying the visual encoding of them \cite{halladjian2022multiscale}
For example, systems can remove visual encodings by removing diverse color attributes in individual pieces of a object such as a genome \cite{halladjian2022multiscale}, pixelate an image into only its key color swatches \cite{shi2023destijl}, or remove text labels that describe properties of generated text to reduce visual fidelity in the interface \cite{suh2024luminate}. 

Design patterns that remove details are overview+detail and focus+context interfaces, which aim to help users gain a bigger picture of their activities \cite{wongsuphasawat2011lifeflow, arendt2020parallel, weninger2020evaluating,  kim2021githru, yoon2015semantic} and data \cite{yu2020skyline, cuenca2018multistream, abello2002mgv, abello2006askgraphview, coscia2024iscore}.
Overview+detail interfaces reduce the amount of information to process by showing only a subset of data attributes of each item in a list \cite{min2025malleable} or by scaling down the view of a map to provide a ``minimap'' to keep the user oriented when navigating the space \cite{cockburn2009review}.
Similarly, focus+context interfaces condense the view of data but provides a magnified, selected view of information in-context of the space \cite{cockburn2009review}.

\paragraph{\textbf{Adding Complexity}}

However, simplifying information is not always the best way to form abstractions. Simplifying can outcast important aspects of information, such as latent patterns about data \cite{chen2014visual, chalmers1996adding} and the inner workings of a system \cite{spolsky2002leakyabstractions}.
To this end, researchers also take the design approach to add more details into the interface, often in ways such that they are not cognitively overwhelming yet reveal richer patterns about the information.

One common technique is to add more details from a previous representation that lacks it. In scatterplots, large numbers of overlapping points often cause ``overdraw'', where dense regions become visually saturated and obscure meaningful structure. One solution is clustering by replacing individual points with a single outline and color \cite{mayorga2013splatterplots}. However, this also removes the ability to perceive patterns within clusters. Chen et al. tackled this trade-off by restoring internal detail without excessive overdraw, enabling users to see both the overall cluster structure and the finer patterns inside \cite{chen2014visual}. These kinds of details can also take the form of visual encodings \cite{chalmers1996adding, jacobs2017supporting}, animations between different visualizations \cite{pu2021datamations}, icons that carry more context than citation numbers \cite{chang2023citesee}, and color tints over UI components to reveal interaction frequency \cite{willett2007scented, hill1992editwearreadwear}.

Although abstraction techniques that add more details in the interface still try to reduce the cognitive load of engaging with the system, there are still many scenarios in which the system to understand is simply too complex, and simplifying would lead to an incomprehensive understanding of the information to communicate.
Thus, various systems embrace the complexity of information and provide abstraction of content that are richer with details and more interactive. These interfaces often aim to match the complexity of the underlying system, allowing users to gain a deeper understanding of it.
Examples of this technique include visual programming languages for authoring full-stack applications \cite{scaffidi2016londontube} and debugging systems for machine learning models \cite{schlegel2020towards, wang2023deepseer}.


Another technique is to increase granularity of information by presenting the individual units altogether, highlighting their commonalities and variations.
For example, recent research has prompted large language models not only to return a single response to a user prompt, but to generate dozens of variations. Such outputs reveal the broader design space for writing and programming opportunities \cite{suh2024luminate, zamfirescupereira2025beyond} and help users more effectively identify the most suitable output \cite{gero2024supporting, wu2023scattershot}.
By presenting a broad range of variations derived from a single instance, the interface supports users in forming their own abstractions of the content.

\paragraph{\textbf{Specificity of Elements}}

Techniques for specifying abstract information commonly involve translating abstract user intents into concrete instructions. As opposed to formalized instructions such as program instructions or buttons on the interface, abstract intents can lack the specificity needed to execute a set of system instructions. This requires design techniques to specify these intents more accurately. One technique involves working towards accurately identifying the user's intent by reverse engineering their modes of expression. This can involve breaking down prompts, gestures, and sketches into various instruction categories and executing them \cite{tilekbay2024expressedit, zhutian2023sporthesia, arora2019magicalhands}. Another technique involves supporting multiple modalities and representations to engage with, such as enabling users to prompt the system with natural language while also making refinements through direct manipulation \cite{kazi2017dreamsketch}, demonstration \cite{li2017sugilite}, or programming \cite{angert2023spellburst, yen2024coladder, tao2025designweaver}.

Freeform input modalities, however, do not necessarily imply unspecified forms of expression. For example, an interactive system may constrain freeform input to enforce concreteness---such as requiring users to speak in Java code rather than natural language prompts \cite{begelspoken}, or supporting notational programming, which allows code to be ``written'' through sketches of notations in context \cite{arawjo2022notational}.




\subsubsection{\textbf{Dimension 3: Representation}}\textit{How does the abstraction take form in the interface?}
\label{section:representation}

This dimension describes how the abstractions are ultimately take form in the interface.
Commonly, interface elements represented into lists, grids, and tables \cite{jain2025plaid, matejka2013swifter, huang2024table}, as well as spatial representations such as node-link canvases, maps, and timelines \cite{hartmann2010dnote, kalnikaite2010where, ge2021cast}.
The abstractions can also be represented into various media, such as images, videos, text documents, sketching spaces, color palettes, conversational interfaces, and 3D scenes for graphic design or AR/VR \cite{l2021dynamic, cheng2009smartplayer, yeh2018craftml, heimerl2016citerivers, mangano2014supporting, shi2023destijl, yeckehzaare2020qmaps, kaplan2025contextural}.
They can also take shape in data-driven representations such as charts and plots (e.g., \cite{pu2021datamations, arendt2020parallel, liao2018clusterbased, kim2013bristle}) and also outputted to the user through different modalities such as audio for blind users \cite{taeb2024axnav, chundury2024tactualplot}.

\paragraph{\textbf{Familiarity and Realism}}

A design approach for providing representations that are more digestible by users is by transforming them into more ``accurate'' or ``realistic'' representations. 
For example, this can be seen when mapping LaTeX components into their rendered forms, enabling more direct and intuitive interaction with LaTeX code  \cite{gobert2022ilatex}, representing SVG code as directly manipulable shapes \cite{hempel2016semiautomated, hempel2019sketchnsketch, zhou2023filteredink}, situating virtual environments with physical reality \cite{hartmann2019realitycheck, rogers2019exploring}, and simplifying the complexities of digital fabrication code with a simulation of the expected result \cite{twiggsmith2024knitscape}.

Systems can also make content and interaction more familiar by utilizing metaphors---by pulling in more familiar elements we interact with, such as using the metaphor of puzzle pieces for authoring data tables \cite{huang2024table}, video keyframes for authoring user interfaces \cite{wu2024framekit}, rubber as a material to afford novel ways to manipulate maps \cite{sarkar1993stretching}, digital art tools for manipulating text \cite{masson2025textoshop}, and web programming languages for similarly programming 3D models \cite{yeh2018craftml}.

\paragraph{\textbf{Creating a Proxy Over a Collection of Elements}}

A technique for effectively representing and interacting with a collection of elements is by creating a proxy that represents the collection. A common form of proxy is a representation that displays the attributes that are common across the content in that collection. For example, a collection of objects on a graphic design tool can be represented by common colors, strokes, or shapes, and selecting and editing an attribute selects all objects associated with that attribute \cite{xia2017collectionobjects, kwon2011direct, shi2024piet, shi2024exploring}.
Another form of proxy selects  a ``representative'' item from the collection, as explored in video navigation systems that intelligently select and present relevant frames when navigating to points on a video \cite{cheng2009smartplayer}.

\paragraph{\textbf{Reifying Operations}}

Another technique directly involves encapsulating system instructions into interactions supported by the interface and reifying them as manipulable objects and tools. A typical example is found in systems that provide higher-level functions over a set of instructions, such as reusable rules for processing data or tools and brushes for rapidly creating custom textures and patterns \cite{jacobs2018extending, kazi2014draco, felice2016beyond, mackay2025interactionsubstrates}.

\subsubsection{\textbf{Dimension 4: Transformability}}\textit{How transformable are the abstractions for the end-user?}
\label{section:transformability}



The transformability of abstractions involves the degree in which end-users can form their own abstractions of information on the interface. While the Presentation dimension describes how systems can provide multiple abstractions for the user to navigate between, these abstractions may be provided by the designers and developers, in which case, constrains the user into toggling between them. Research also explored enabling end-users to develop their own abstractions with the generative constraints provided by the system designers and developers.


\paragraph{\textbf{Fixed Single Abstraction}}
Most systems provide fixed abstractions with the goal of introducing a new abstraction level that enables users to understand and interact with information in novel ways. With this in mind, research systems will aim to focus on this contribution before considering how their novel abstractions can be transformable. For example, The Object-Oriented Drawing systems proposed Attribute Objects to enable users to more directly interact digital attributes \cite{xia2016objectoriented}. Draco \cite{kazi2014draco} and Kitty \cite{kazi2014kitty} introduced kinetic textures and motion amplifiers, respectively, to enable users to create animated illustration with ease. 





\paragraph{\textbf{Predefined Sets of Abstractions}}

A number of interactive systems propose presenting information with multiple abstractions, with the purpose of providing multiple perspectives into specific problems and content and highlighting the connections between them \cite{victor2011ladder}. Abstractions presented in these interfaces involve transformations across levels of details or multiple types of schemas and representations.
Interfaces that provide multiple levels of detail may, for example, generate progressively more concise summaries of text \cite{suh2023sensecape, mirowski2023cowriting}, incrementally unravel genome sequences \cite{halladjian2022multiscale, halladjian2020scale}, remove data and visual attributes in interfaces representing structured data \cite{choi2025genpara, suh2024luminate, min2025malleable}. Interfaces that provide multiple schemas and representations to provide more distinct perspectives. For instance, a system may guide students from reasoning about the physical world to reasoning about mathematics or code \cite{suh2022codetoon, saquib2021constructing}, or enable users to shift from editing individual pixels of digital ink to manipulating strokes, and ultimately to composing shapes and structures \cite{xia2017writlarge}. Systems providing multiple abstractions need to also incorporate corresponding interface mechanisms to enable users to view, navigate, and synchronize content among the different abstractions as we will discuss in the next dimension.


\paragraph{\textbf{Generative and Malleable Abstractions}}

Beyond offering predefined sets of abstractions, systems can support generative and malleable abstractions by equipping end users with low-level creation media such as drawing and programming media \cite{landay2002sketching, kay1977personal}, high-level composable components \cite{xia2016objectoriented, xia2017collectionobjects, xia2018dataink, min2025malleable} and generative rules expressed through domain-specific languages \cite{cao2025generative, hayatpur2025shapes}. For example, Attribute Objects can form aggregated graphical styles \cite{xia2016objectoriented}, selection filters \cite{xia2017collectionobjects}, and visual encodings \cite{xia2018dataink}. Overviews in an overview+detail interface are abstractions of the details, and can be made transformable by allowing end-users to select which data attributes to be shown or hidden in the overview \cite{min2025malleable}. Using a DSL that describes the abstraction and transformation moves of data-structure diagrams, Chisel allows flexible specification and transformation of such diagrams \cite{hayatpur2025shapes}. By integrating LLMs with generative, task-driven schemas and configurable interface mappings, Jelly enables the generation and transformation of user interfaces tailored to individual users’ needs and goals \cite{cao2025generative}.

\subsubsection{\textbf{Dimension 5: Presentation}}\textit{How does the interface situate and display abstractions?}
\label{section:presentation}


This dimensions concerns how abstractions of content and operations are presented in the interface relative to surrounding system components. These design choices influences how users perceive, interpret, and interact with the abstractions, as well as other information in the interface.

\paragraph{\textbf{Replacing the Original}}

As covered in previous dimensions, designing abstractions involves transforming a previous abstraction with its representation into a new one. One way to present these transformations is by replacing the previous representation to enable and encourage users to interact with this new form, 
such as new environments for authoring content \cite{vlissides1991unidrawbased, ge2021cast, zhang2022onelabeler}, new visualizations for data \cite{heimerl2016citerivers, arendt2020parallel}, and control interfaces that allows for higher-level forms of expression \cite{kim2022stylette, barbosa2018zenstates}.
Many systems that replace a previous representation also provide ways to switch back or even toggle between multiple representations \cite{sun2014role, mangano2014supporting, diehl2021hornero}.

\paragraph{\textbf{Presenting, Navigating, and Synchronizing Multiple Abstractions}}

When multiple abstractions are incorporated in the system, interface mechanisms need to be provided to enable users to easily view and navigate among the abstractions.  Typically, this is done by presenting different representations side-by-side (i.e., juxtaposed \cite{lobo2015evaluation}) \cite{subramonyam2020texsketch, latif2019author, hammond2010creating, hedderich2024piece} or in a gallery/dashboard arrangement \cite{assogba2011many, maram2024ah, beck2024choreovis, shin2025drillboards}.
There are also various approaches to presenting the abstractions more in-situ of the rest of the content, such as inserting abstractions in-line \cite{cheng2024relic, methfessel2024mse,latif2019author, coscia2024iscore, zou2025gistvis} or interleaved between lines of text \cite{cheng2024relic, gero2024supporting}, inserting them into a 2D/3D scene \cite{pongnumkul2008creating, muresan2023using, zhou2024timetunnel, obrenovic2011sketching}, blending or overlaying the abstraction into the view \cite{lobo2015evaluation, wallner2019aggregated, satriadi2020maps, l2021dynamic}, or animating changes in the transformations \cite{pu2021datamations, masson2023charagraph}. Other novel interface mechanisms for navigating multiple abstractions include semantic zooming \cite{bederson1994pad}, space folding \cite{elmqvist2010mlange}, and representational axes \cite{xia2017writlarge}.

When multiple abstractions are simultaneously represented on the interface, synchronization mechanism is often required to ensure changes of content within one representation is synchronized with the others \cite{chevalier2012histomages, koch2014varifocalreader, cao2025compositional, suh2022codetoon}, such as via synchronized highlighting \cite{cao2025compositional}, editing \cite{xia2020crosspower, contentbasedediting}, and animated transitions \cite{textanimation}.


\subsubsection{\textbf{Dimension 6: Guidance}}\textit{How does the interface influence and scaffold the user’s workflow with the provided abstractions?}
\label{section:guidance}




The dimension of guidance describes design techniques that control which abstractions users encounter and engage with in the interface. This influences how effectively users engage with the information, which can adhere to a recommended workflow of completing a task or implementing a design guideline on the order of representations to show.




\paragraph{\textbf{Organizing Which Abstractions to Interact With}}

One approach to guiding users through abstraction in the interface is by organizing which abstractions they first encounter and which ones they interact with later in the task.
The most notable technique used in interaction design is Shneiderman's Mantra: ``Overview First, Zoom and Filter, Details on Demand'' \cite{bederson2003shneidermanmantra}. This guideline has been widely adopted across data visualization systems and user interfaces for decades.
However, in recent years, academics have begun to propose opposing views of this guideline for different contexts. Luciani et al. argued for: ``Details First, Show Context, Overview Last'' instead to support domain experts who already have a rich overview of data \cite{luciani2019detailsfirst}.
There continues to be a discussion and search for which mantra is suitable for which contexts.

\paragraph{\textbf{Transitioning Abstractions Throughout a Task}}

Another approach involves more scaffolding in the interface and are commonly implemented for learning and understanding purposes.
Concreteness Fading is a notable pedagogy design technique, where the user is taught an abstract concept, such as mathematics or code, by first being introduced familiar and concrete representations (e.g., oranges and apples), then gradually decreasing the level of concreteness towards the abstract representation \cite{bruner1966concretenessfading, suh2020concretenessfadingsurvey}.
This form of scaffolding has been instantiated mostly in programming learning systems \cite{weinman2021improving, suh2022codetoon, arawjo2017teaching, yeckehzaare2020qmaps}.

This approach of revealing more complexity about a knowledge source throughout a task has also been explored in diverse contexts, such as bridging an understanding gap between AI engineers and designers to collaborate more effectively \cite{subramonyam2022solving}, increasing the level of complexity of quiz questions \cite{oberdo2019usability}, or facilitating the development of professional vision towards the effects of color filters \cite{beaudouinlafon2023color} and object outlines \cite{l2021dynamic}.









%% file: sections/6_abstraction_model.tex

\section{Abstraction Spaces}

Our design space illustrates that abstractions in both the interactive system and the user’s mental model shape the dynamics of interaction. However, the Gulfs of Execution and Evaluation do not explicitly model how these abstractions influence interaction dynamics between the user and system.
Abstraction Spaces, an interaction model for abstraction, reframes the Gulfs of Execution and Evaluation to explicitly represent abstraction.

Our Abstraction Spaces interaction model is based on several key findings from the survey.
First, abstraction designs do not influence only a single stage of Norman’s seven stages of action—they shape all stages (Sec. \ref{section:design_goals}). This suggests that introducing an orthogonal dimension of abstraction across the Gulfs may reveal richer patterns of how abstraction designs influence interaction.
Second, we observe a parallel in how abstractions can be performed by either the user or the system. For instance, a user may read through a large body of text and abstract it into key points, or the system may perform the same abstraction on the user’s behalf by summarizing the text.
Third, many challenges in abstraction stem from mismatches in abstraction level. Existing systems often provide abstractions that are either too high-level or too low-level for the designer’s goals. Designing the ``right'' abstraction involves either aligning the system’s abstraction level with the user’s needs or deliberately misaligning them to encourage the user to view a perspective.
Finally, although systems typically provide single abstraction layers and abstraction ``ladders,'' we find that the abstractions in a user’s mental model extend well beyond these constraints. This suggests that we must consider the design space of abstraction as a space that spans a much broader range than most systems currently support.

Abstraction Spaces adds the dimension of abstraction between the system and user to the traditional Gulfs model. This involves three components:
\begin{enumerate}
\item The \textbf{\textit{Gulf of Abstraction}} adds to the two traditional Gulfs.
\item \textbf{\textit{Abstraction Overshoots and Undershoots}}, adds two categories of abstraction misalignment.
\item The \textbf{\textit{Space of Abstractions}}, illustrates how the gulf of abstraction spans not only a single vertical axis, but a broader space.
\end{enumerate}

\begin{figure*}
    \centering 
    \includegraphics[width=1\linewidth]{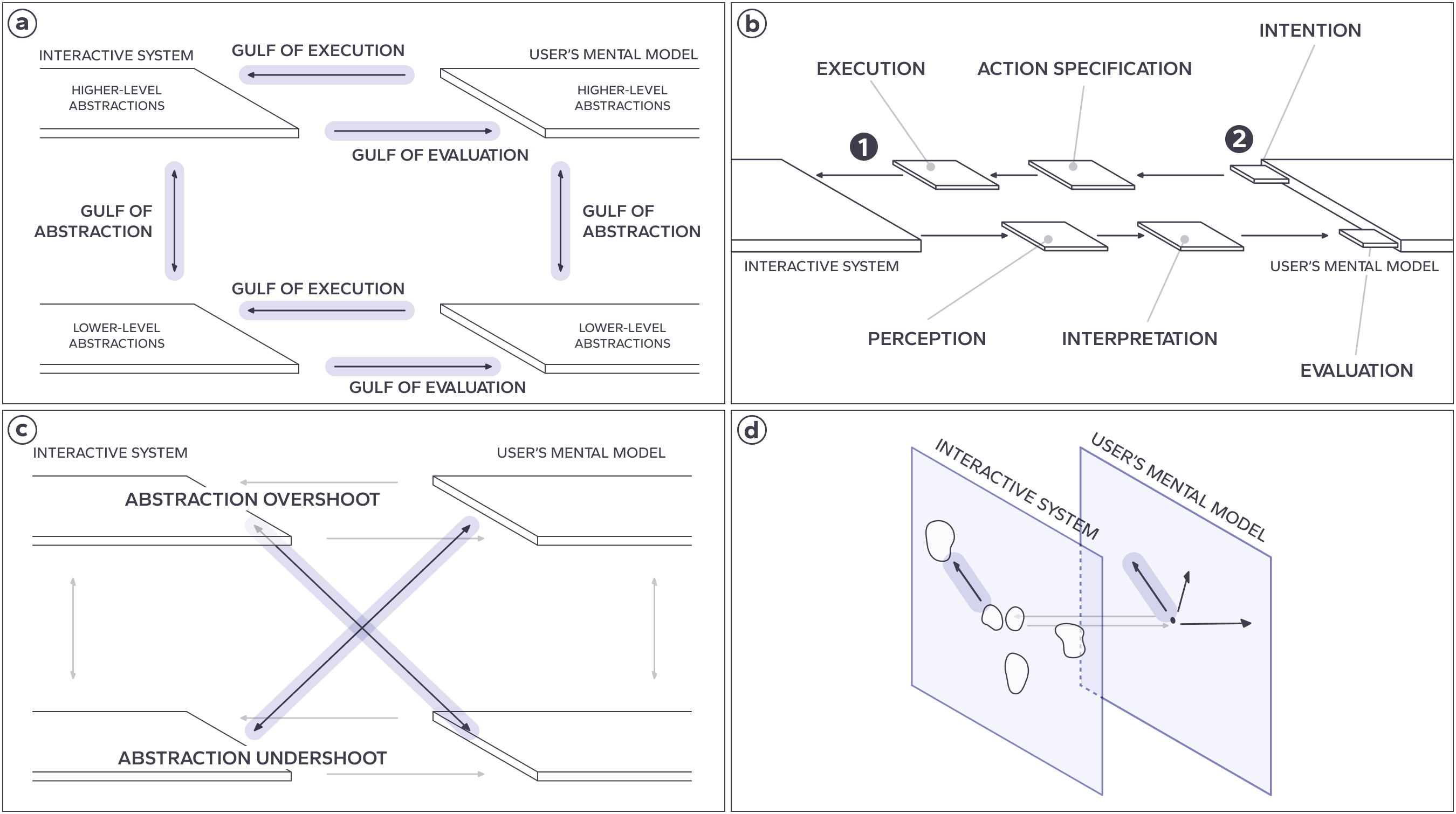}
    \caption{(a) The interaction model of abstraction re-contextualizes the traditional Gulfs of Execution and Evaluation with an added Gulf of Abstraction. (b1) The stages of action in the Gulfs of Execution and Evaluation are involve the four stages after the user's mental model, as depicted as well in (Fig. \ref{fig:design-purpose-stages}). (b2) The stages of intention and evaluation are involved in the user's cognitive process and reside in the Gulf of Abstraction. (c) Abstraction undershoot and overshoot influence how the system or user must span the abstraction gulf to form alignment. (d) The space of abstractions aims to illustrate the opportunity of considering a broader space of abstractions in designing interactive systems.}
    \label{fig:abstraction-model-components}
    \Description{Components of the Abstraction Spaces model. (a) Extension of Norman’s gulfs with a vertical Gulf of Abstraction. (b) Mapping of Norman’s stages of action onto abstraction shifts. (c) Illustration of abstraction overshoots and undershoots between system and user models. (d) 3D schematic showing abstraction alignment and misalignment across planes of the interactive system and user’s mental model. }
\end{figure*}

\subsection{The Gulfs of Execution, Evaluation, and Abstraction}

In order to make abstraction explicit in the Gulfs interaction model, we first introduce the Gulf of Abstraction, and then redefine the traditional Gulfs of Execution and Evaluation.


\paragraph{\textbf{Gulf of Abstraction}}

The Gulf of Abstraction refers to the misalignment between the user’s mental model and the system’s representation of information. This distance can be bridged when the two models become aligned---either by the user adapting their mental model to the system’s abstractions, the system adapting its abstractions to better match the user’s needs, or both.

Spanning the Gulf of Abstraction involves two complementary processes: (1) On the user’s side, the activity includes \textit{intent formation} and \textit{evaluation}, as users translate their goals into the system’s available abstractions and assess whether the system’s outputs align with their expectations. (2) On the system’s side, the activity involves \textit{abstraction transformation techniques} described in the design space, such as adjusting the granularity, the specificity of information or user expressions, and the scope of information transformed.
In our interaction model, the Gulf of Abstraction is primarily represented as a vertical axis (Fig.~\ref{fig:abstraction-model-components}a), highlighting the distance between higher- and lower-level abstractions that must be traversed for effective interaction.

\paragraph{\textbf{Gulfs of Execution and Evaluation}}

In Norman’s framework \cite{norman1988psychology}, the Gulfs of Execution and Evaluation map onto stages of the seven stages of action where users transform intentions into executable actions and evaluate the outputs they perceive from the system. Specifically, the ``intention'' stage marks where users translate their goals into actionable intents for the system, while the ``evaluation'' stage marks where users compare their interpretation of the system’s output against their goals.
By contrast, spanning the \textit{Gulf of Abstraction} involves both the intention and evaluation stages. Forming an intention requires selecting the level of abstraction that best aligns with the user’s goal, while evaluation requires abstracting the system’s output in order to assess its correspondence to that goal.

We therefore reframe the Gulfs of Execution and Evaluation: rather than encompassing the \textit{processes} of intention formation and outcome evaluation, they instead denote the \textit{end points} of their respective gulfs. This distinction is illustrated in Figure~\ref{fig:abstraction-model-components}b.





\subsection{Abstraction Overshoots and Undershoots}

This model surfaces two categories of abstraction misalignments: overshoots and undershoots (Fig~\ref{fig:abstraction-model-components}c).

\paragraph{\textbf{Abstraction Overshoots}}

Abstraction overshoots occur when the system provides a higher level of abstraction than the user’s mental model, or when the user shifts to a lower level of abstraction than the system provides. As a result, users need to exert additional cognitive effort to align their mental model with the system before crossing the Gulfs of Execution and Evaluation. For example, an abstraction overshoot occurs when an interface presents an overly concise summary of text, forcing the user to dig through source documents to find the answers they need. Another example is when a user instructs a system using voice commands, but cannot steer it precisely enough using this modality. At first, the user may align with the system's level of abstraction in expressing high level intents through voice commands, but they may reach cases where these intents are not recognized accurately.


\paragraph{\textbf{Abstraction Undershoots}}

Abstraction undershoots occur when the system provides a lower level of abstraction than the user’s mental model, or when the user shifts to a higher level of abstraction than the system provides. In such cases, the interface may present information that is too detailed or complex, requiring the user to mentally process and construct abstractions on their own. Undershoots can also arise as users become more experienced with a system. For example, an expert Photoshop user may find its tools too granular, as they must repeatedly recreate similarly styled objects using only individual brush strokes.

\subsection{Space of Abstractions}


Although we primarily illustrate the Abstraction Spaces model as a vertical axis, the design space of abstractions in interactive systems is not linear—it spans a broader \textit{space of abstractions}. For example, given a text document, a designer might abstract it by summarizing the contents into a single paragraph. Yet such a summary could take many different forms: it might emphasize the narrative flow, describe the figures, or extract the references. Users, in turn, have diverse needs for different kinds of information, and a single summary cannot capture the full range of possible abstractions.

To highlight this broader space, we illustrate how the user’s mental model and the system can span the gulf of abstraction not only by shifting to higher or lower levels, but by moving through a space (Fig.~\ref{fig:abstraction-model-components}d). The solutions for bridging the gulf remain the same: the system can shift toward the user, the user can translate their mental model toward the system, or both can shift toward one another and meet at a middle ground.







\section{Examining Interaction Using Abstraction Spaces}

Abstraction Spaces derives from the broad literature of HCI interactive systems designing for abstraction. Thus, we are interested in how our model can model existing workflows and perspectives.





















\subsection{Modeling Existing Workflows}

We aim to show how Abstraction Spaces explicitly models interaction processes relating to the design of abstractions that the original Gulfs model does not.
To demonstrate this, we examine (1) how our model stays consistent with the original gulfs framework while modeling abstraction more explicitly and (2) how our model can describe interaction dynamics of existing novel interactive systems in HCI.
We analyze four workflows across different interaction dynamics and model them using both the original Gulfs framework as well as ours.

\begin{figure*}
    \centering
    \includegraphics[width=1\linewidth]{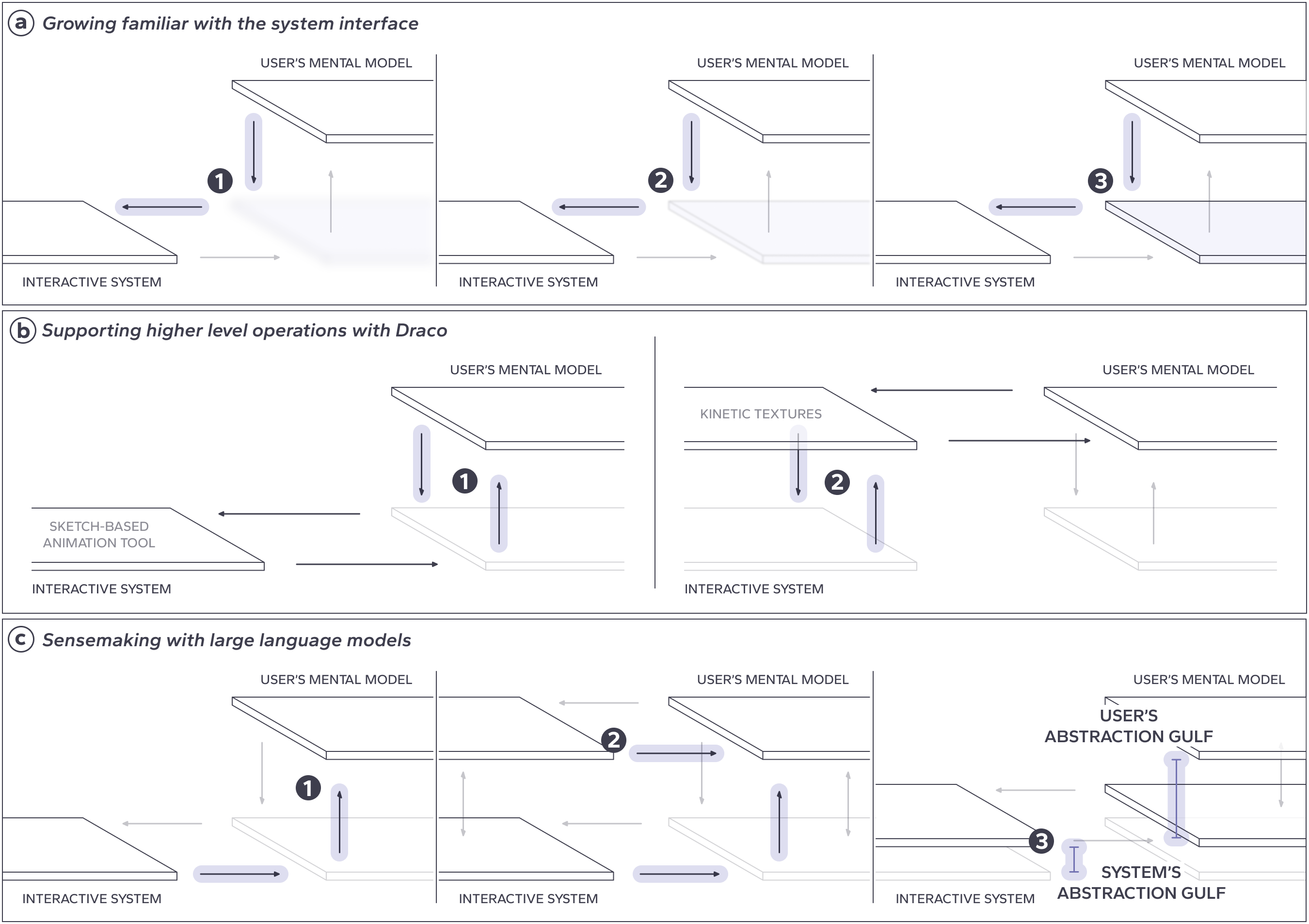}
    \caption{(a) As the user grows familiar with the system's level of abstraction, they are able to span the abstraction gulf easier. (b) Systems such as Draco \cite{kazi2014draco} can enable users to interact at higher levels of abstraction, aligning closer to their desired sequence of actions. (c) Systems that support sensemaking of AI-generated information can help users by either forming abstractions for the user such as full text summaries (1) or form an abstraction only to a certain extent, such as highlight key phrases, and leave the rest for the user to synthesize (2).}
    \label{fig:modeling-workflows}
    \Description{Examples of applying the Abstraction Spaces model to interaction workflows. (a) Growing familiar with a system interface as the user’s abstractions align with the system. (b) Supporting higher-level operations with Draco by shifting abstractions across tools. (c) Sensemaking with large language models, highlighting mismatches between user and system abstraction gulfs. }
\end{figure*}

\paragraph{\textbf{Automating Behavior}}

As tasks are practiced frequently, they become more familiar to the user and require less cognitive effort to complete. Hutchins et al. note that such automated behavior does not reduce the semantic distance but rather users overcome this distance through continual use and practice \cite{hutchins1986directmanipulation}. This semantic distance corresponds to the Gulf of Abstraction and can be illustrated as different levels of abstraction in the user’s mental model (Fig. \ref{fig:modeling-workflows}a).
Following Hutchins et al.’s example of gaining familiarity with the vi text editor, a novice user may struggle to span this abstraction gulf when mapping a high-level intent—such as deleting a word—to the command, ``dw’’ (Fig. \ref{fig:modeling-workflows}a.1), because they have not formed a robust mental approximation of the system abstraction. However, as they repeatedly locate and execute the commands, they begin to form more accurate and stable approximation of the system and become more efficient in moving from their abstraction of the tasks to the abstraction of the system  (Fig. \ref{fig:modeling-workflows}a.2–3). We can illustrate this familiarity in our interaction model by gradually fading in the user’s abstraction level as it aligns with the system.

\paragraph{\textbf{Reifying Operations}}

In the workflow of users automating their own behavior, they often encounter systems that are not aligned with the abstractions in their mental model. In such cases, even experienced users may find actions tedious despite their familiarity with the interactions. To illustrate this, we examine the research system Draco \cite{kazi2014draco}, a tool designed to reduce tedious, repetitive strokes for authoring sketch-based animations.
Traditional sketch-based animation tools require users to manually create textures or elements to animate within a scene. However, animating large collections in this way can be tedious, as users must repeatedly apply similar animations and textures to each element. While a user may already hold a higher-level abstraction of the operations they wish to perform—namely, animating the collection of elements as a whole—traditional systems constrain them to operate at a lower, more granular level (Fig. \ref{fig:modeling-workflows}b.1).
Draco addresses this gap by aligning with the user’s higher-level abstractions. It does so by forming abstractions over the process of creating textures and reifying them into brushes that users can readily apply. The strokes produced with these brushes then animate the textures along the user’s drawn paths.
The Abstraction Spaces model captures this workflow and design approach as a case where the system forms abstractions that mirror those already present in the user’s mental model  (Fig. \ref{fig:modeling-workflows}b.2).

\paragraph{\textbf{Sensemaking with Large Language Models}}

Sensemaking is a fundamental activity in interacting with information. The sensemaking process involves searching, understanding, organizing, schematizing, refining, and presenting information \cite{russellsensemaking}. The emergence of LLMs has enabled ways to significantly reduce the costs involved in the sensemaking stages.
One line of work leverages LLMs to search and collect information, enabling users to focus on understanding the information.
However, since LLMs are capable of rapidly generating large amounts of text, collecting and generating text is not a bottleneck, but understanding it efficiently and deeply remains the challenge to overcome.

Abstraction Spaces models this problem space as an abstraction undershoot, where the user must form abstractions of large amounts of LLM-generated text to understand its key points (Fig \ref{fig:modeling-workflows}c.1). Our design space highlights two approaches to this problem: the system can form abstractions of the text for the user, or the user can form these abstractions on their own.
The system can form abstractions by summarizing the text, enabling users to view the relationship between the original text and the summarized one \cite{suh2023sensecape}. This can be modeled by the system spanning the abstraction gulf (Fig. \ref{fig:modeling-workflows}c.2).
On the other hand, the system can leave the user to form abstractions on their own. Though, the system can still provide abstractions that assist with this process, such as highlighting common phrases across different bodies of text to help users identify connections easier \cite{gero2024supporting}). This can be modeled by the system spanning the abstraction gulf only partially, leaving the user to span the rest of the gulf on their own (Fig. \ref{fig:modeling-workflows}c.3).

\begin{figure*}
    \centering
    \includegraphics[width=1\linewidth]{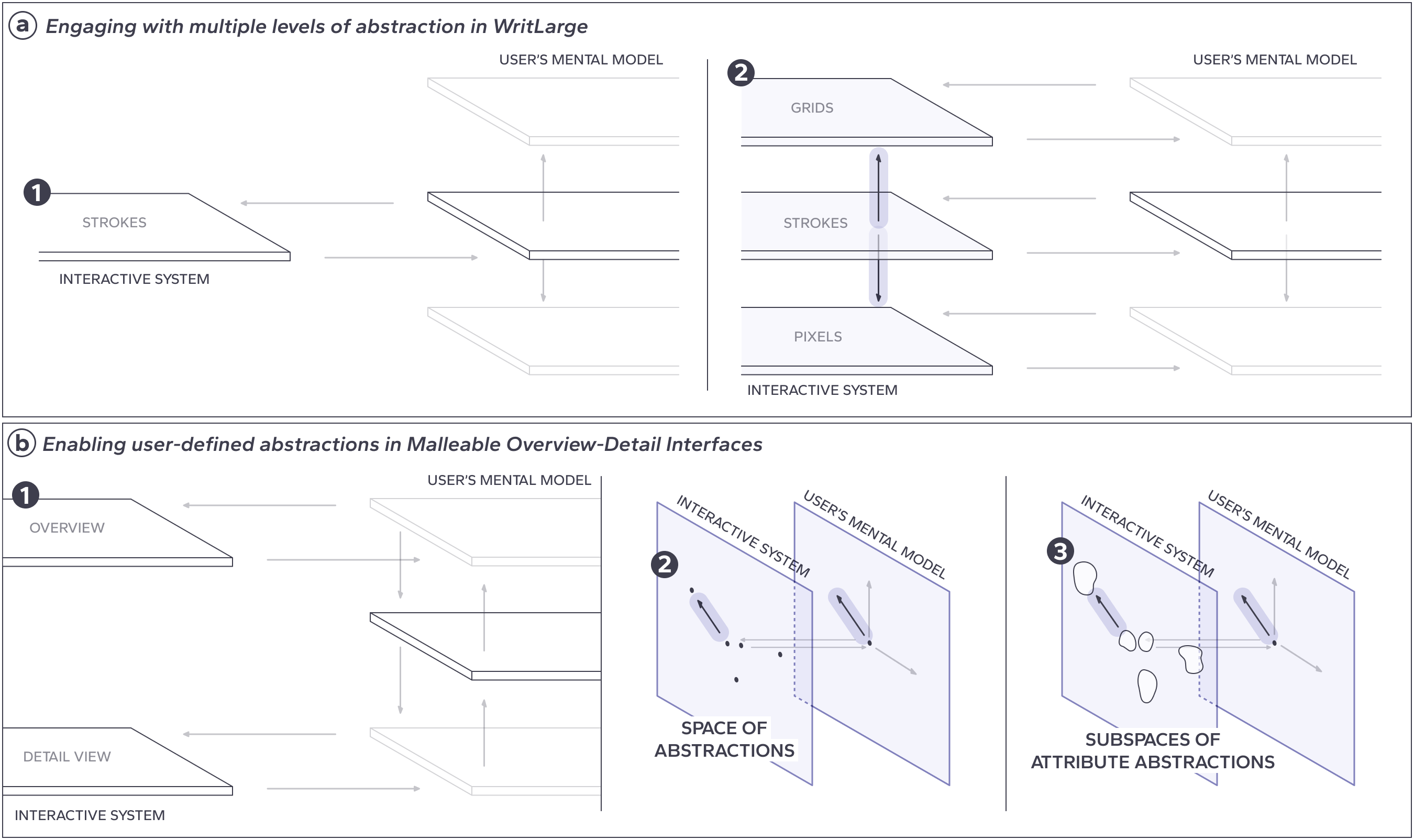}
    \caption{(a) Systems can provide many levels of abstraction for users to engage with. WritLarge \cite{xia2017writlarge} enables users to transform a sketched grid into a grid component itself, or down to the individual pixels of each stroke, giving users the expressiveness to manipulate the systems' information in flexible ways. (2) Systems can also provide various abstractions that are not constrained into a single ladder of abstraction. Malleable overview-detail interfaces \cite{min2025malleable} enables users to customize the abstractions they wish to see in the interface by transforming representations and the set of attributes shown on the interface.}
    \label{fig:modeling-workflows-2}
    \Description{Applying the Abstraction Spaces model to real systems. (a) Engaging with multiple levels of abstraction in WritLarge, from strokes to grids to pixels. (b) Enabling user-defined abstractions in malleable overview–detail interfaces, illustrating how users span abstraction spaces and subspaces of attribute abstractions. }
\end{figure*}

\paragraph{\textbf{Engaging with Multiple Levels of Abstraction}}


Research on engaging with multiple levels of abstraction highlights how users can view and interact with information in diverse ways. WritLarge enables such multi-level abstractions to support brainstorming activities on a digital whiteboard \cite{xia2017writlarge}. It addresses a limitation of digital whiteboards that digital strokes are treated as the sole medium of expression. As a result, when a user wants to make more granular, pixel-by-pixel edits to a sketched grid or broader adjustments to the grid as a whole, they are constrained by the stroke-based medium (Fig. \ref{fig:modeling-workflows-2}a.1). WritLarge overcomes this by providing multiple abstractions over the medium of digital elements. Users can transform strokes into pixels for fine-grained modifications and then move back up to adjust the broader grid structure.

We can model WritLarge’s interaction dynamics with these multi-level abstractions by representing the system across distinct levels (Fig. \ref{fig:modeling-workflows-2}a.2). The user’s mental model can then be depicted as moving fluidly across these levels of abstraction, engaging with the whiteboard through their desired abstraction.

\paragraph{\textbf{Forming Abstractions through Malleable Interfaces}}

Many research systems have recently demonstrated the benefits of providing multiple representations of information—whether through different arrangements, modalities, or levels of abstraction. These representations allow users to view information from different perspectives and engage with those that resonate most with their needs. One prominent example is the overview+detail interface, where the overview provides a broad view of many items by omitting most of their attributes, while the detail view reveals the full set of attributes for a selected item \cite{min2025malleable, cockburn2009review}.
However, users’ information needs in overview-detail interfaces are highly diverse. Some users may prioritize one set of attributes, while others focus on entirely different ones \cite{min2025malleable, williams2023data}. To find the key information they need, users must often shuttle back and forth between the overview and detail view, piecing together their information needs. In other words, they are forced to operate across levels of abstraction that do not match their mental model, requiring additional cognitive effort to gather key information (Fig. \ref{fig:modeling-workflows-2}b.1).

To address this, Min et al. \cite{min2025malleable} provides \textit{malleable} overview-detail interfaces, which allow users to customize the kinds of attributes that appears in the overview and in which representation. Examples of representations include list, table, map, or a spatial canvas. We model this flexibility in two ways. First, choosing the representation of the overview is mapped as points in a space of abstractions. Second, customizing shown attributes means users can explore a subspace of possible abstractions within each representation. This means each point expands to become individual subspaces (Fig. \ref{fig:modeling-workflows-2}b.2).

\subsection{Modeling Existing Perspectives}


We also examine existing interaction models to understand how Abstraction Spaces can represent various design spaces related to abstraction. In particular, we analyze Instrumental Interaction and the Gulf of Envisioning, highlighting which components of these models can be effectively captured and which cannot, and what opportunities our model may surface.


\begin{figure*}
    \centering
    \includegraphics[width=1\linewidth]{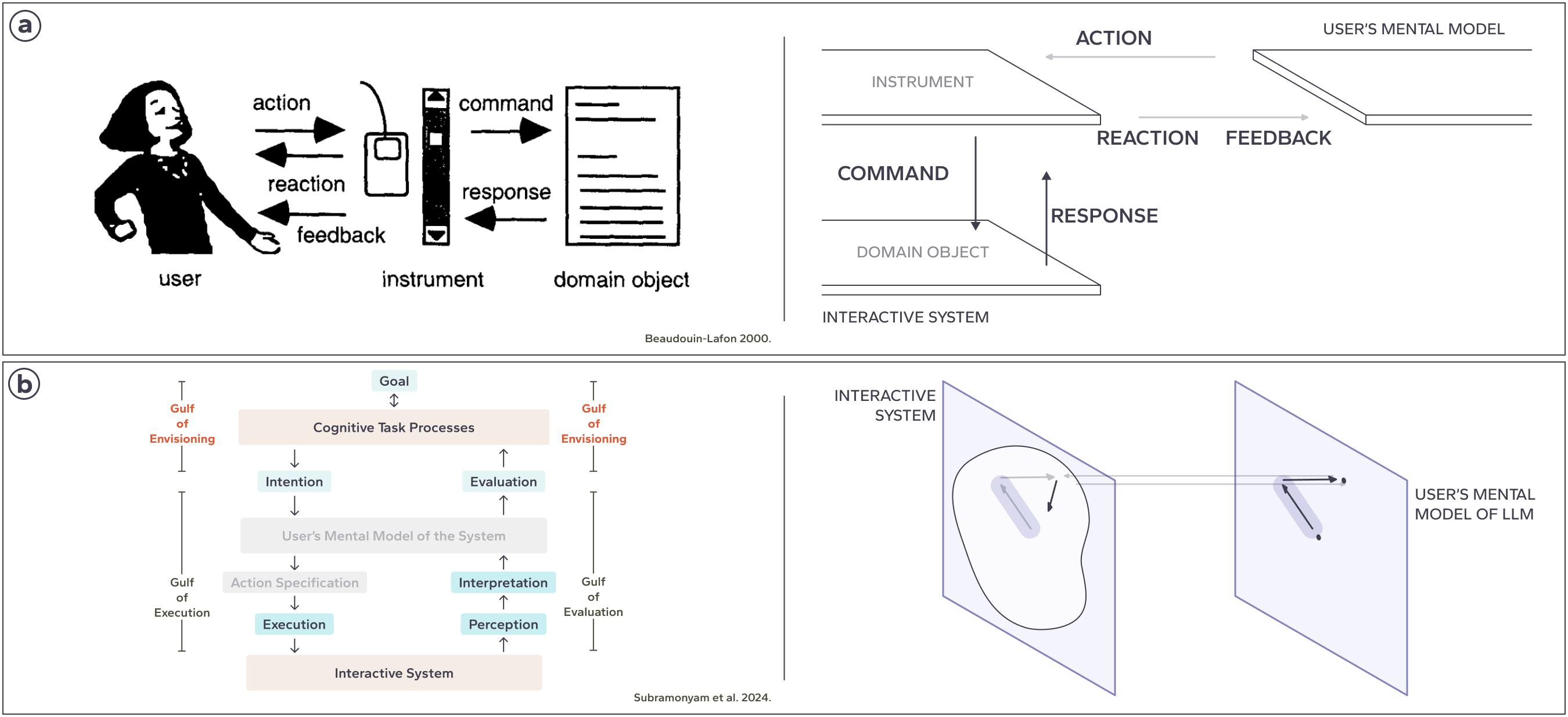}
    \caption{(a) Instrumental Interaction \cite{mbl2000instrumentalinteraction} can be modeled by depicting the instrument at a higher level of abstraction over the system’s provided features on the domain object. This illustrates how instruments enable users to engage with the system through an abstraction layer that is both aligned with the user’s mental model and expressive. (b) The Gulf of Envisioning \cite{hari2024gulfenvisioning} describes the distance between the user’s intentions for engaging with an LLM and the prompt they form. In our model, we can depict this by illustrating how LLMs do not operate at a fixed level of abstraction. As a result, the user must anticipate the LLM’s response and iteratively prompt to achieve alignment.}
    \label{fig:modeling-perspectives}
    \Description{Comparing perspectives on interaction models. (a) Instrumental Interaction showing the roles of user, instrument, and domain object. (b) Extensions of Norman’s gulfs, including the Gulf of Envisioning, illustrated for interactions with large language models.}
\end{figure*}

\paragraph{\textbf{Instrumental Interaction}}

Instrumental Interaction is an interaction model that extends the principles of direct manipulation to the activity of tool use in interfaces \cite{mbl2000instrumentalinteraction}. It introduces instruments, which mediate interaction between the user and system information. Designing such instruments involves three principles: (1) \textit{reifying} a set of operations or properties into manipulable objects, (2) ensuring \textit{polymorphism}, or the breadth of contexts in which the tool can be applied, and (3) identifying the \textit{reuse} of system operations. For example, the alignment functionality in graphics editors, which visualizes objects snapping to each other, can be reified into a manipulable instrument that allows users to create persistent alignments and arrangements of objects \cite{felice2016beyond}.

In Abstraction Spaces, reification is an abstraction technique which involves encapsulating sequences of operations and representing them as manipulable objects on the interface. Through this lens, the model of Instrumental Interaction can be translated into Abstraction Spaces as a process of the system spanning the abstraction gulf through reification and mediating interaction between the user and system (Fig. \ref{fig:modeling-perspectives}a). 
Abstraction Spaces further emphasizes the need for users to have power over the interface, enabling them to appropriate system tools and define their own abstractions \cite{li2023beyond}.

\paragraph{\textbf{Gulf of Envisioning}}

LLMs can respond to user prompts expressed at virtually any level of abstraction. While this enables an extremely dynamic mode of interaction, the responses does not always align with the user’s needs. Consequently, crafting and articulating prompts that yield the desired outcomes with desired abstraction remains challenging in LLM-powered interaction systems. Subramonyam et al. describe this challenge as the gulf of envisioning, which denotes the distance between the user’s intentions and the prompts they construct \cite{hari2024gulfenvisioning}. Users bridge this gulf as they reason about the LLM’s capabilities, decide on prompting strategies, and anticipate how the model might respond.

With our interaction model, as users draw on prior experiences with LLMs, they attempt to speculate how to prompt them to operate with the desired abstraction. However, unlike conventional systems with fixed abstractions, which allows users to form stable abstraction of the system through use, LLM's unpredictable generations hinder the users from forming similarly robust abstraction with the LLM. This is depicted as the user's struggle to align with the LLMs whose abstraction is essentially a moving target.

%% file: sections/9_discussion.tex
\section{Discussion}
\label{section:discussion}

\subsection{Have We Made Abstraction Concrete?}

We aim to enable HCI researchers to engage with the concept of abstraction more explicitly and concretely. To this end, we conducted a large-scale analysis of research that introduces novel design and interaction techniques for abstraction and used our findings to develop an interaction model that places abstraction at its core. As part of this process, we also created a graphical representation of the model, which proved instrumental for depicting and analyzing interactive systems through the lens of abstraction. This representation clearly communicates key dynamics—including the misalignment between system and mental abstractions, the formation of users’ mental abstractions of the system through use, the benefits of offering multiple abstractions, and the expansive space of abstraction afforded by large language models. In effect, this representation has helped us view abstraction in a more concrete and actionable manner. 

Yet our interaction model is not without limitations. First, our analysis has primarily focused on the relationship between system abstractions and users’ mental abstractions, as well as the various dimensions for designing system abstractions. By contrast, the concrete actions associated with users’ own abstraction gulf are less fully represented in the current model. We see was of extension. For example, familiarity with a system can enhance the perceived directness of interaction \cite{hutchins1986directmanipulation}. This could be represented by depicting lower levels of the abstraction gulf as “submerged,” indicating that certain parts of the cognitive process have become automated or performed unconsciously.  Second, although the visual metaphor developed for our interaction model appears promising, its boundaries remain untested. We are particularly interested in conducting broader evaluations of this metaphor in practice, especially when applying it to a wider range of systems.







\subsection{How Do We Design AI to Align or Misalign?}

An important aspect our interaction model highlights is the alignment of abstractions between the user’s mental model and the system. With the rapid rise of generative AI, research has begun to explore diverse approaches to designing human-AI interactions that augment, rather than replace or undermine, human cognition. Yet this remains an open challenge to uncover how we can design interactions in which the AI aligns with the user when appropriate, but also deliberately misaligns in ways that encourage the user to realign and reflect?

We have seen various approaches to this, including designing AI to become less agreeable by providing contrasting ideas from the user's \cite{vaithilingam2024imagining}. Another is to incentivize metacognition in the user through various design strategies that invoke user's awareness of their own prompting practices \cite{tankelevitch2024metacognitive}.

One way that leverages the design of abstraction is by incorporating variations into the generated response \cite{gero2024supporting}. For example, the system might highlight general, unopinionated patterns across these variations—such as common or unique phrases—and encourage the user to form their own abstractions from them. This principle could be extended by exploring other abstraction techniques that achieve a similar effect. We might model this interaction and identify design techniques that follow this dynamic. For instance, an LLM could be designed to exaggerate its behaviors—such as deliberately responding in an overly verbose manner—prompting the user to revise their prompt for a more succinct output. Such nudges can help users more readily notice when the system’s behavior does not align with their goals and preferences, potentially bridging the Gulf of Envisioning in prompt formulation.

Another approach is to design feedforward mechanisms for users as they prompt AI \cite{min2025feedforward}. In the context of LLMs, feedforward can involve providing previews of how the model might respond before the user submits their prompt. These previews may take the form of meta-data about the anticipated response—such as its length, the types of representations it might include, or even a brief inner dialogue revealing what the model is planning to generate \cite{liu2025innerthoughts}.

An alternative design strategy is to have the AI infer the user’s higher-level goals from their prompt. In this view, the system abstracts from the prompt to construct a model of the user’s goals.
Recent research has begun to explore the generation of user models from individuals’ daily activities.
Building on such models, one way to either align or deliberately misalign with the user is to treat the constructed model as contextual ground truth. The system may then generate two responses: one that adheres to the user model’s context and another that intentionally diverges from it. Presenting both aligned and misaligned responses may support users in reflecting on, evaluating, and refining their goals.

Taken together, these approaches illustrate how the deliberate design of abstractions can create opportunities for designing human-AI systems that align or misalign with the user, thereby augmenting and preserving their cognition and control over the interface.

\subsection{From Singular, to Ladder, to Space}
One of the initial motivations for our work was a sense of confinement imposed by the concept of the Ladder of Abstraction, despite its clear advancement over most existing systems \cite{victor2011ladder}. We found that this ladder metaphor enforces a rigid dimensionality of abstraction—progressing from low-level to high-level. By contrast, the HCI literature increasingly presents systems that support multiple, non-hierarchical abstractions that resist reduction to a single low-to-high continuum \cite{cao2025compositional, min2025malleable}. Moreover, framing abstraction exclusively along this continuum encourages extending existing abstractions only through encapsulation or decomposition. Yet truly novel abstractions may depart from this axis entirely, offering orthogonal perspectives that generate fundamentally new insights. Our current visual metaphor—a plane representing the space of abstraction—also feels constraining. Ultimately, we aim to encourage the research community to think beyond the ladder and to explore abstractions that unlock genuinely new perspectives.







%% file: sections/99_conclusion.tex
\section{Conclusion}
\label{section:conclusion}


Abstraction shapes every facet of interaction, yet HCI lacks an interaction model that explicitly accounts for how abstraction influences the user’s cognitive process of interaction. Drawing on our survey of 457 papers across HCI venues, we developed a design space of abstraction techniques spanning six dimensions.
Building on this analysis, we propose an interaction model—Abstraction Spaces—that reframes the traditional Gulfs of Execution and Evaluation by incorporating a new Gulf of Abstraction. We then examined how this model can describe diverse interaction workflows in existing systems and perspectives. Through Abstraction Spaces, we aim to provide researchers with a lens for analyzing systems through abstraction and to encourage them to imagine how to design abstractions beyond a single layer or ladder.

%% file: sections/999_appendix.tex
\clearpage

\section{Appendix}
\label{section:appendix}

\subsection{Exclusion Criteria for the Survey}
\label{appendix:criteria}

\textbf{1. Does not contribute an interactive system.}

We excluded empirical studies, surveys, frameworks, and technical/performance contributions.
We excluded empirical studies because our focus is how researchers' interpretation of ``abstraction'' maps to design components of interactive systems rather than behavioral findings.
We excluded theory, framework, and survey papers since they do not describe the design of systems themselves, although we did include papers in which a framework was instantiated as an interactive system.
Lastly, we excluded papers that contributed algorithms or models aimed primarily at improving the technical performance of an interactive system (e.g., recommender system, constraint solver, fine-tuned AI model).

\textbf{2. Uses ``abstraction'' in an irrelevant context.}
As we examined the papers, we found many that mentioned the term ``abstraction'' in the paper and contributed an interactive system, yet were still out of our scope because they did used the term in an \textit{irrelevant context}.
We determined that ``abstraction'' was used in an irrelevant context when the term:

\begin{itemize}
    \item Described the cognitive process or phenomena of a real-life activity (e.g., learning, sensemaking, writing).
    \item Described the activity of synthesizing concepts from a qualitative analysis or literature review task.
    \item Described the design of a conceptual model or internal data model of a system (e.g. hardware/software architecture)
    \item Described the property of a grammar or language that formalizes a domain (e.g., programming library).
    \item Described a feature in an interactive system of a different paper (this was common in related work sections).
    \item Was mentioned by user study participants rather than paper authors themselves.
    \item Referred to the first section of academic papers (i.e., the ``Abstract'')
    \item Appeared in parts of the paper unrelated to its research content (e.g., references section, copyright statement).
\end{itemize}